\documentclass{aa}
\usepackage{graphicx}
\usepackage{txfonts}
\begin{document}
\title{
Imaging the outward motions of 
clumpy dust clouds around the red supergiant Antares with VLT/VISIR\thanks{
Based on VISIR observations made with the Very Large Telescope of 
the European Southern Observatory. Program ID: 385.D-0120(A), 286.D-5007(A)}
}
%\subtitle{
%}

\author{K.~Ohnaka
}

\offprints{K.~Ohnaka}

\institute{
Max-Planck-Institut f\"{u}r Radioastronomie, 
Auf dem H\"{u}gel 69, 53121 Bonn, Germany\\
\email{kohnaka@mpifr.de}
}

\date{Received / Accepted }

\abstract
{}
% Aim
{
We present a 0\farcs5-resolution 17.7~\mbox{$\mu$m}\ image of the red supergiant 
Antares.  Our aim is to study the structure of the circumstellar envelope 
in detail.  
}
% Methods
{
Antares was observed at 17.7~\mbox{$\mu$m}\ with the VLT mid-infrared instrument 
VISIR.  
Taking advantage of the BURST mode, in which a large number of short exposure 
frames are taken, we obtained a diffraction-limited image with a spatial 
resolution of 0\farcs5.  
}
% Results
{
The VISIR image shows six clumpy dust clouds located at 0\farcs8--1\farcs8 
(43--96~\mbox{$R_{\star}$}\ = 136--306~AU) away from the star.  We also detected 
compact emission within a radius of 0\farcs5 from the star.  
Comparison of our VISIR 
image taken in 2010 and the 20.8~\mbox{$\mu$m}\ image taken in 1998 with the 
Keck telescope reveals the outward motions of four dust clumps.  The proper 
motions of these dust clumps (with respect to the central star) 
amount to 0\farcs2--0\farcs6 in 12 years.  
This translates into expansion velocities (projected onto the plane 
of the sky) of 13--40~\mbox{km s$^{-1}$}\ with an uncertainty of $\pm7$~\mbox{km s$^{-1}$}. 
The inner compact emission seen in 
the 2010 VISIR image is presumably newly formed dust, because it is not 
detected in the image taken in 1998.  If we assume that the dust is ejected 
in 1998, the expansion velocity is estimated to be 34~\mbox{km s$^{-1}$}, in agreement 
with the velocity of the outward motions of the clumpy dust clouds. 
The mass of the dust clouds is estimated to be $(3-6)\times10^{-9}$~\mbox{$M_{\sun}$}. 
These values are lower by a factor of 3--7 than the amount of dust ejected 
in one year estimated from the (gas+dust) mass-loss rate of 
$2\times10^{-6}$~\mbox{$M_{\sun}$~yr$^{-1}$}, 
suggesting that the continuous mass loss is 
superimposed on the clumpy dust cloud ejection. 
}
% Conclusions
{
The clumpy dust envelope detected in the 17.7~\mbox{$\mu$m}\ diffraction-limited 
image is similar to the clumpy or asymmetric circumstellar environment of 
other red supergiants. 
The velocities of the dust clumps cannot be explained by a simple accelerating 
outflow, implying the possible random nature of the dust cloud ejection 
mechanism. 
}

\keywords{
infrared: stars --
techniques: high angular resolution --
stars: supergiants  --
stars: late-type --
stars: mass loss --
stars: individual: Antares
}   %  END OF ABSTRACT

%\titlerunning{}
\authorrunning{Ohnaka}
\maketitle

\section{Introduction}
\label{sect_intro}

Mass loss is important for understanding the evolution of 
massive stars.  In the red supergiant (RSG) phase, massive stars experience 
intense mass loss.  The RSG mass loss significantly affects the evolution of 
massive stars, and it is a key to understanding the progenitors of 
core-collapse supernovae.  Nevertheless, the mass-loss mechanism in the RSG 
phase is a long-standing problem, and the driving force of the RSG mass loss 
has not been identified yet.  

Recent high spatial resolution observations of RSGs have revealed 
complex asymmetric structures in the region close to the star. 
The near-IR imaging of the optically bright RSGs \object{Betelgeuse} 
(\object{$\alpha$~Ori}) and \object{Antares} (\object{$\alpha$~Sco}) 
shows asymmetric and clumpy structures (Cruzal\`ebes et al. 
\cite{cruzalebes98}; Kervella et al. \cite{kervella09}).  
In particular, the images 
of Betelgeuse taken at 1.04--2.17~\mbox{$\mu$m}\ with spatial resolutions of 
27--56~mas by Kervella et al. (\cite{kervella09}) 
shows a plume extending to $\sim$130~mas (= $\sim$6~\mbox{$R_{\star}$}). 
Ohnaka et al. (\cite{ohnaka09}; \cite{ohnaka11}; \cite{ohnaka13}) carried out 
high spatial and high spectral resolution observations of 
Betelgeuse and Antares in the CO first overtone lines near 2.3~\mbox{$\mu$m}\ 
using the near-IR interferometric instrument AMBER at the Very Large Telescope 
Interferometer (VLTI).  
Their ``velocity-resolved'' aperture-synthesis images 
revealed temporally variable, inhomogeneous gas motions in the photosphere 
and the molecular outer atmosphere (so-called MOLsphere) 
extending to $\sim$1.5~\mbox{$R_{\star}$}. 
The detected motions are 
qualitatively similar to the motions of the hotter chromospheric gas 
spatially resolved by Lobel \& Dupree (\cite{lobel01}).  
These observations indicate that the material is not spilling out in an 
ordered, spherical fashion. 

Asymmetric, inhomogeneous structures are also found on larger spatial scales. 
The near-IR imaging of the dusty RSGs \object{VY~CMa} and 
\object{NML~Cyg} suggests bipolar outflows 
and/or equatorial disks with even more complex fine structures 
(e.g, Wittkowski et al. \cite{wittkowski98}; Kastner \& Weintraub 
\cite{kastner98}; 
Monnier et al. \cite{monnier04}; Humphreys et al. \cite{humphreys07}).  
Noticeable deviation from spherical symmetry is 
revealed even in an RSG in an extragalactic system: 
Ohnaka et al. (\cite{ohnaka08}) spatially resolved the torus 
around the dusty RSG \object{WOH~G64} 
in the Large Magellanic Cloud using the mid-IR 
interferometric instrument MIDI at VLTI.  
Non-spherical mass loss is detected in optically bright 
(i.e., not very dusty) RSGs as well.  
The mid-IR imaging of Betelgeuse and Antares 
by Hinz et al. (\cite{hinz98}), Kervella et al. (\cite{kervella11}), and 
Marsh et al. (\cite{marsh01}) revealed asymmetric and/or clumpy circumstellar 
environment extending up to $\sim$100~\mbox{$R_{\star}$}.  
De Wit et al. (\cite{dewit08}) detected elongation in the circumstellar 
envelope of \object{$\mu$~Cep} at 24.5~\mbox{$\mu$m}, 
which they interpret as the possible evidence of a slowly expanding torus.  
The 24.5~\mbox{$\mu$m}\ 1-D intensity profiles of Antares and \object{$\alpha$~Her} 
presented by de Wit et al. (\cite{dewit09}) also show extended circumstellar 
envelopes, although they do not discuss asymmetry. 
The inhomogeneous gas motions detected in the outer atmosphere might be 
the seed of the asymmetric and/or clumpy structures seen in the 
circumstellar envelope.

In this paper, we present 0\farcs5-resolution, diffraction-limited 
mid-IR imaging of the circumstellar envelope of Antares at 17.7~\mbox{$\mu$m}\ 
with VLT/VISIR.  
We also report on the detection of the outward motions of clumpy dust clouds 
over 12 years.  
Antares (M1.5Iab-b) is a well-studied prototypical RSG at a distance of 170~pc 
(based on the parallax from van Leeuwen \cite{vanleeuwen07}) 
with a moderate mass-loss rate of 
$\sim \!\! 2 \times 10^{-6}$~\mbox{$M_{\sun}$~yr$^{-1}$}\ (Braun et al. \cite{braun12}). 
From its effective temperature ($3660\pm120$~K) and luminosity 
($\log \mbox{$L_{\star}$}/\mbox{$L_{\sun}$} = 4.88\pm0.23$), its mass is estimated to be 
$15\pm5$~\mbox{$M_{\sun}$}\ (Ohnaka et al. \cite{ohnaka13}).  
Antares has a hot companion (B2.5V) at a separation of 2\farcs7, which 
can be used to probe the mass loss from the primary RSG 
(e.g., Baade \& Reimers \cite{baade07}; Reimers et al. \cite{reimers08}).

\section{Observations and data reduction}
\label{sect_obs}

\begin{table}
\begin{center}
\caption {Summary of the VISIR observations.  
NDIT: Number of frames. Seeing is in the visible.  
}

\begin{tabular}{l l l r c c}\hline
 Object & UTC & DIT  & NDIT &  seeing  & airmass \\
        &     & (ms) &      &  (\arcsec) &   \\
\hline
\multicolumn{6}{c}{2010 June 02}\\
\hline
Antares & 01:42:59 & 12.5 & 24000 & 1.2 & 1.26 \\
        & 02:09:07 & 20.0 & 12000 & 1.2 & 1.17 \\
        & 02:44:39 & 20.0 & 12000 & 1.2 & 1.09 \\
        & 03:08:37 & 20.0 & 12000 & 1.2 & 1.05 \\
        & 03:31:05 & 20.0 & 12000 & 1.5 & 1.03 \\
        & 03:58:25 & 20.0 & 36000 & --- & 1.01 \\
\mbox{$\varepsilon$~Sco} & 03:41:43 & 20.0 & 24000 & 1.5 & 1.05 \\
\mbox{$\lambda$~Sgr} & 02:57:01 & 20.0 & 12000 & 1.2 & 1.48 \\
        & 03:19:55 & 20.0 & 12000 & 1.2 & 1.35 \\
\hline
\multicolumn{6}{c}{2010 November 12 (archived data)}\\
\hline
Aldebaran&07:11:03 & 12.5 & 10240 & 1.3 & 1.42 \\
\hline
\label{obslog}

\end{tabular}
\end{center}
\end{table}

\subsection{VISIR observations}
\label{subsect_obs_obs}

We observed Antares with VLT/VISIR (Lagage et al. \cite{lagage04}) 
on 2010 June 2 (UTC) using the Q1 filter 
centered at 17.7~\mbox{$\mu$m}\ with a half-band width of 0.83~\mbox{$\mu$m}.  
VISIR is equipped with a 256$\times$256 BIB detector with pixel scales of 
0\farcs075 and 0\farcs127.  We used the pixel scale of 0\farcs075 for 
our observations of Antares.  
The observations were carried out with chopping and nodding to subtract 
the sky background.  
We used a chopping and nodding angle of 8\arcsec\ with 
the direction of the chopping and nodding set to be perpendicular.  
This results in four images on the detector 
after processing the chopped and nodded frames. 
The chopping frequency was 0.5~Hz, and the nodding period was 90~sec. 

We took advantage of BURST mode (Doucet et al. \cite{doucet07}), which takes 
a number of exposures with a short detector integration time (DIT) to freeze 
the atmospheric turbulence.  
This allows us to obtain a diffraction-limited image, 
which is difficult to achieve in normal long exposures (see, e.g., 
Kervella \& Domiciano de Souza \cite{kervella07} for comparison of the 
images taken in BURST mode and usual long-exposure mode).  
As Table~\ref{obslog} summarizes, 
we took 108000 frames for Antares and 24000 frames for the 
calibrators \object{\mbox{$\varepsilon$~Sco}} and 
\object{\mbox{$\lambda$~Sgr}}, using DITs of 12.5 and 20~ms. 

We observed these calibrators not only for the flux calibration of the Antares 
image but also as references of the point spread function (PSF).  
However, 
as we present below, while the Antares image shows 
up to the tenth Airy ring, 
we detected only up to the third Airy ring in the image 
of \mbox{$\varepsilon$~Sco}\ and only the central core in the image of \mbox{$\lambda$~Sgr}, 
because these calibrators are much fainter than 
Antares\footnote{No brighter calibrators could be observed because the 
telescope had to be closed due to strong winds.}. 
This makes the quality of the PSF-subtracted image of Antares 
remarkably worse than before the PSF subtraction. 
Therefore, as a second PSF reference, 
we downloaded archived VISIR BURST mode imaging data of \object{Aldebaran} 
(\object{$\alpha$~Tau}) taken on 2010 November 12 with the Q1 filter 
(Program ID: 286.D-5007A, published in Kervella et al. \cite{kervella11}) 
and reduced them in the same manner as our data.  

\subsection{Data reduction}
\label{subsect_obs_reduction}

\begin{figure*}
%\resizebox{\hsize}{!}{\rotatebox{0}{\includegraphics{alfsco_destriping_4x1.ps}}}
\resizebox{\hsize}{!}{\rotatebox{0}{\includegraphics{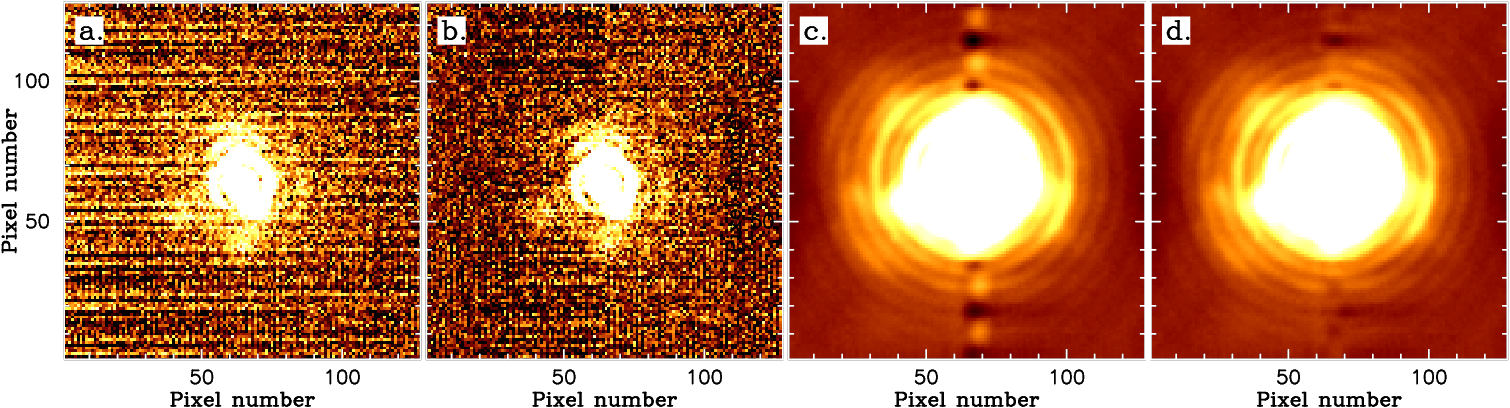}}}
\caption{
Removal of the detector artifacts.  
{\bf a:} One of the chop-nod processed images of Antares, showing 
horizontal stripes. 
{\bf b:} The same image as in the panel {\bf a}, after the removal of the 
horizontal stripes, as described in Sect.~\ref{subsect_obs_reduction}. 
{\bf c:} Shift-and-added image of Antares obtained from all frames, in which 
the horizontal stripes are removed.  The image shows an artifact that 
appears as the vertical regular pattern, whose amplitude is $\sim$0.2\% of 
the peak intensity of the central star. 
{\bf d:} The same image as in the panel {\bf c}, after removing the 
vertical pattern, as described in Sect.~\ref{subsect_obs_reduction}. 
}
\label{alfsco_destriping}
\end{figure*}

The data reduction of the BURST mode data is as follows. 
We first removed the sky background by subtracting the chopped and 
nodded images.  Since the chopping and nodding direction are 
perpendicular to each other, we obtain four images after this 
procedure.  However, in the data of Antares, one of the four images falls 
onto a region significantly affected by a group of bad pixels, and 
it must be discarded.  
In addition, as Fig.~\ref{alfsco_destriping}a shows, 
the images after the chopping and nodding subtraction 
show noticeable horizontal stripes, which are reported by Kervella et al. 
(\cite{kervella11}).  The horizontal stripes are not fixed to specific 
rows but appear in different rows in each frame.  
To remove these detector artifacts, we applied the 
following method presented by Kervella et al. (\cite{kervella11}) to each 
image after the chopping and nodding subtraction. 
At each row, we computed the median in 20 pixels from the left (right) 
edge of the image and subtracted it from the left (right) half of the 
image.  This procedure mostly removes the horizontal stripes, 
as Fig.~\ref{alfsco_destriping}b shows.  
Then we recentered each image and added all images to obtain the 
final images of Antares and the calibrators with 
IRAF\footnote{IRAF is distributed by the National Optical Astronomy
  Observatory, which is operated by the Association of Universities for
  Research in Astronomy (AURA) under cooperative agreement with the National
  Science Foundation.}.

However, the shift-and-added images still show some residual of the horizontal 
stripes, which appears as a regular vertical pattern in columns near the 
center (Fig.~\ref{alfsco_destriping}c).  
The amplitude of the vertical pattern is $\sim$0.2\% of the peak intensity 
of the central star in case of Antares.  
While this appears to be small, we attempted 
to remove the artifact to minimize its effects on the study of the faint 
circumstellar structures.  The vertical pattern appeared in the 
shift-and-added images of Antares and Aldebaran but not in the images of 
the calibrators \mbox{$\varepsilon$~Sco}\ and \mbox{$\lambda$~Sgr}, which are much fainter than the 
former two stars.  To remove this artifact, we fitted 
it with a sinusoidal curve outside the region dominated by the bright 
central core of the image.  Specifically, to remove only the artifact while 
leaving the Airy pattern intact, we first estimated the Airy pattern in the 
affected central 15 columns by interpolating from the adjacent pixels and 
subtracted the interpolated Airy pattern in each column.  
The remaining artifact was fitted with a sinusoidal function in each column 
in the region outside the bright Airy rings (outside the fifth and third Airy 
rings for Antares and Aldebaran, respectively. See also 
Figs.~\ref{alfsco_res1} and \ref{alfsco_res2}c).  The fitted 
sinusoidal pattern was subtracted for all pixels (i.e., also for pixels 
excluded from the sinusoidal fitting) in each column.  
The resulting image (Fig.~\ref{alfsco_destriping}d) shows that the vertical 
pattern is mostly removed.

\subsection{Flux calibration}
\label{subsect_obs_fluxcal}

We carried out the flux calibration of the image of Antares using 
the calibrators \mbox{$\varepsilon$~Sco}\ and \mbox{$\lambda$~Sgr}.  
We adopted the Q1-band flux of 18.59~Jy and 9.9~Jy for \mbox{$\varepsilon$~Sco}\ and 
\mbox{$\lambda$~Sgr}, respectively, taken from the catalog of the mid-IR standard stars on 
the VISIR website\footnote{http://www.eso.org/sci/facilities/paranal/instruments/visir/tools/\\zerop\_cohen\_Jy.txt}.  
The 17.7~\mbox{$\mu$m}\ flux of Antares calibrated with \mbox{$\varepsilon$~Sco}\ and \mbox{$\lambda$~Sgr}\ 
is 1135~Jy and 1282~Jy, respectively.  
With \mbox{$\lambda$~Sgr}\ much fainter than \mbox{$\varepsilon$~Sco}, the 
image quality of \mbox{$\lambda$~Sgr}\ is very poor, making the flux obtained with 
this calibrator less reliable than obtained with \mbox{$\varepsilon$~Sco}.  Therefore, 
we take the value derived with \mbox{$\varepsilon$~Sco}\ as the flux of Antares and 
the difference in the flux derived with \mbox{$\varepsilon$~Sco}\ and \mbox{$\lambda$~Sgr}\ as the 
uncertainty in the flux calibration ($1135\pm148$~Jy).  
The flux derived from the spectrum obtained with the Infrared 
Space Observatory (TDT number: 08200369) and the response function 
of the Q1 filter\footnote{http://www.eso.org/sci/facilities/paranal/instruments/visir/inst/\\Q1\_rebin.txt} is 1224~Jy, 
which agrees with the $1135\pm148$~Jy obtained above within the error.

\section{Results}
\label{sect_res}

\begin{figure*}
\sidecaption
\resizebox{12cm}{!}{\rotatebox{0}{\includegraphics{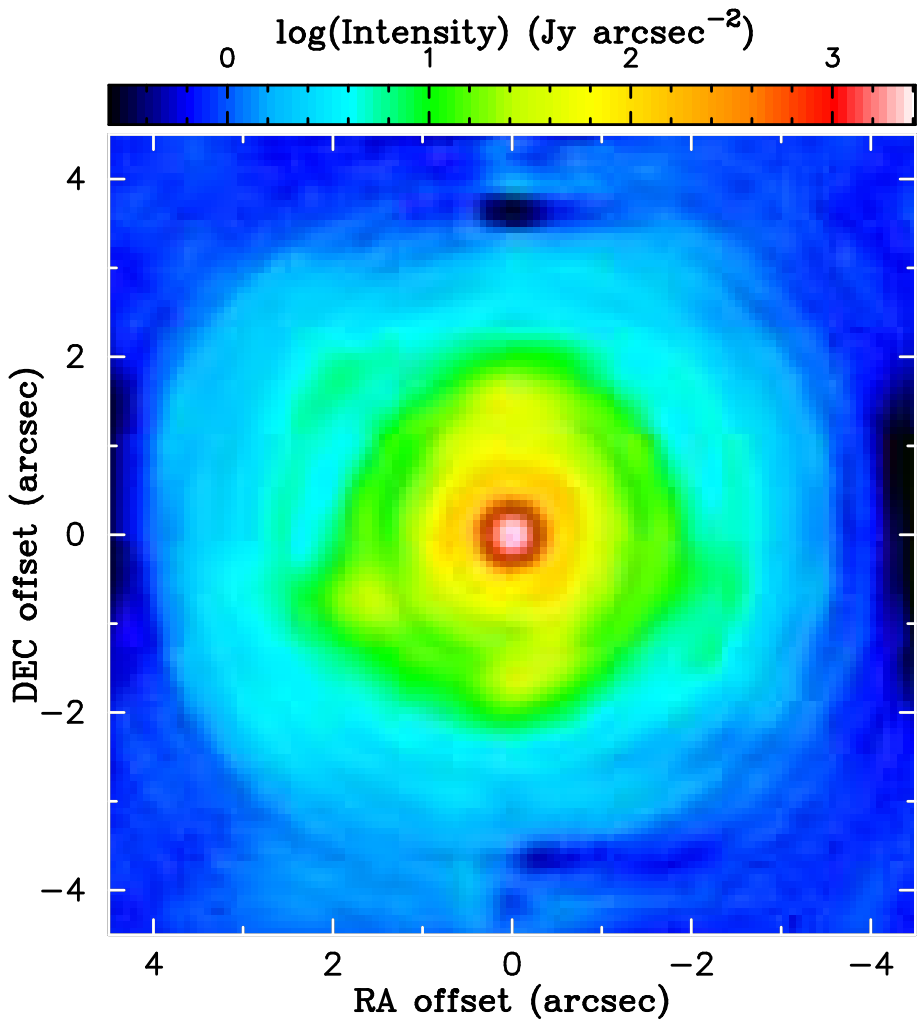}}}
%\resizebox{12cm}{!}{\rotatebox{0}{\includegraphics{alfsco_res1.ps}}}
%\resizebox{\hsize}{!}{\rotatebox{0}{\includegraphics{alfsco_res1.ps}}}
\caption{
Flux-calibrated 17.7~\mbox{$\mu$m}\ image of Antares. The colors are shown on a 
logarithmic scale.  North is up, and east to the left. 
}
\label{alfsco_res1}
\end{figure*}

\begin{figure}
\resizebox{\hsize}{!}{\rotatebox{-90}{\includegraphics{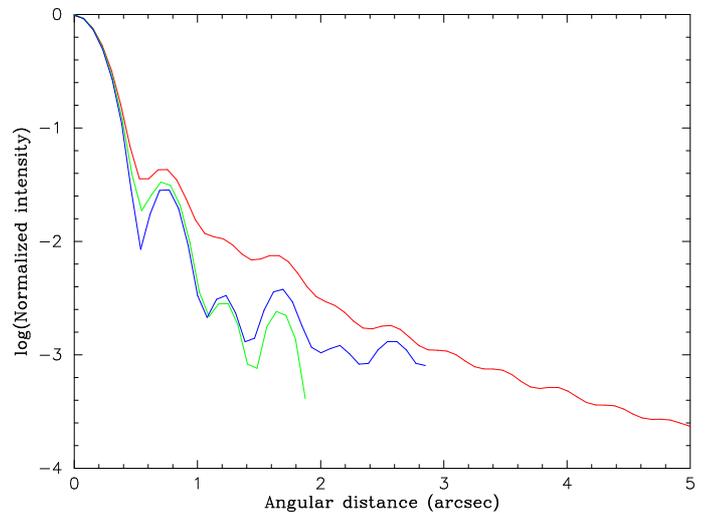}}}
%\resizebox{\hsize}{!}{\rotatebox{-90}{\includegraphics{alfsco_img1Davg.ps}}}
\caption{
Azimuthally averaged intensity profiles of Antares (red), 
\mbox{$\varepsilon$~Sco}\ (green), and Aldebaran (blue).  The intensity profiles are 
normalized with the peak intensity. 
}
\label{alfsco_img1Davg}
\end{figure}

\begin{figure*}
\sidecaption
\resizebox{12cm}{!}{\rotatebox{0}{\includegraphics{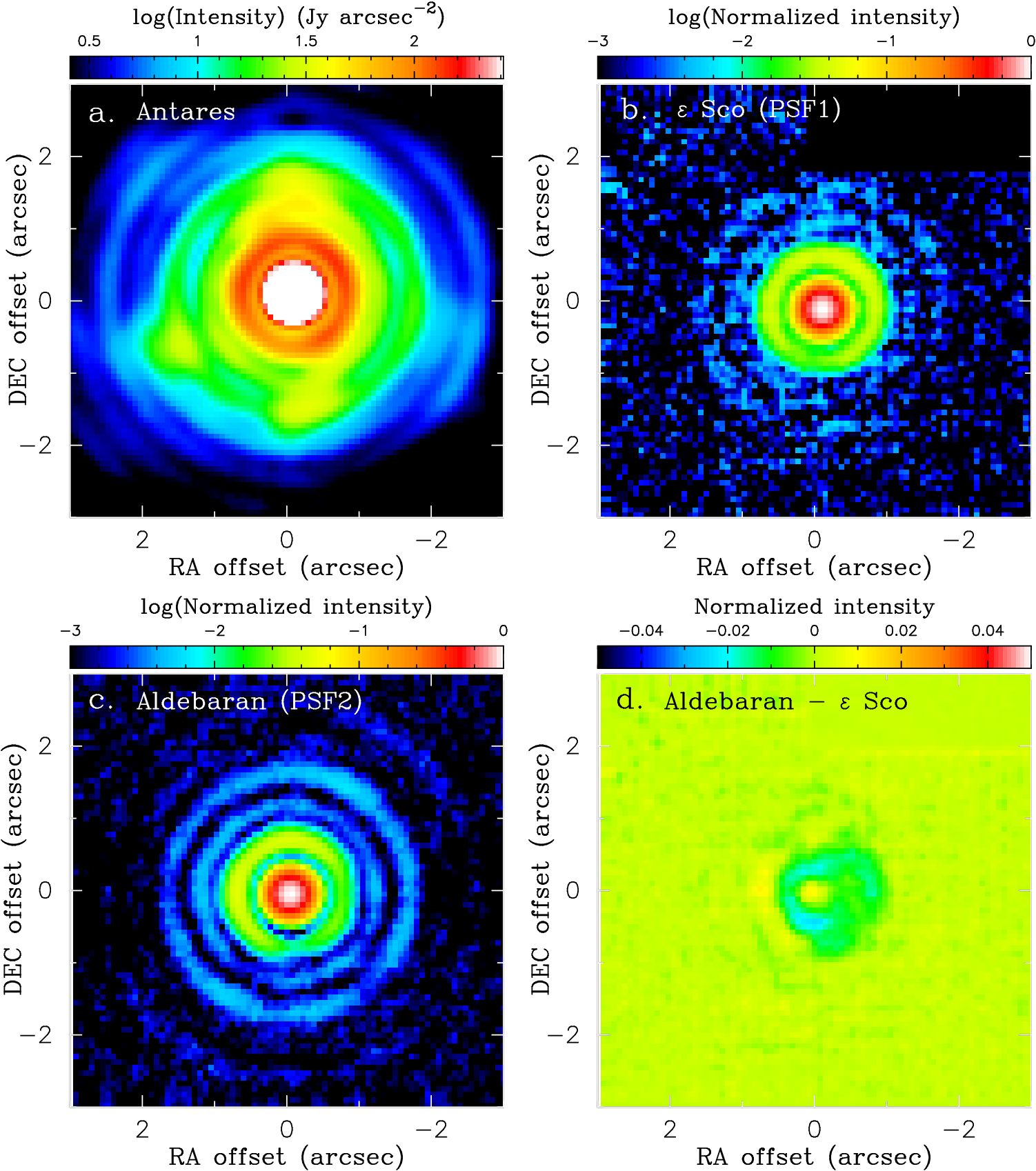}}}
%\resizebox{12cm}{!}{\rotatebox{0}{\includegraphics{alfsco_res2.ps}}}
%\resizebox{\hsize}{!}{\rotatebox{0}{\includegraphics{alfsco_res2.ps}}}
\caption{
Enlarged view of the inner 3\arcsec $\times$\,3\arcsec\ region of the 
flux-calibrated Antares image, together with the PSF references \mbox{$\varepsilon$~Sco}\ 
and Aldebaran.  North is up, and east to the left. 
{\bf a:} Flux-calibrated image of Antares.  The colors are shown on a 
logarithmic scale, but the colors in the central region are saturated. 
{\bf b:} Image of the PSF reference \mbox{$\varepsilon$~Sco}.  The intensity is normalized 
with the peak intensity.  The colors are shown on a logarithmic scale. 
{\bf c:} Image of the PSF reference Aldebaran, shown in the same manner 
as in the panel {\bf b}. 
{\bf d:} Difference between the images of \mbox{$\varepsilon$~Sco}\ and Aldebaran.  
The colors are shown on a linear scale, ranging from $-5$\% to 
5\% of the peak intensity of the PSF reference images. 
}
\label{alfsco_res2}
\end{figure*}

Figure~\ref{alfsco_res1} shows the flux-calibrated 17.7~\mbox{$\mu$m}\ image of 
Antares. 
The high brightness of Antares and the BURST mode allowed 
us to detect up to the tenth Airy ring, which is clearly seen in the 
azimuthally averaged intensity profile shown in Fig.~\ref{alfsco_img1Davg}.  
Comparison of the intensity profile of Antares with those of \mbox{$\varepsilon$~Sco}\ and 
Aldebaran reveals the extended circumstellar envelope.  
We also show 
an enlarged view of the inner 3\arcsec $\times$\,3\arcsec\ region of the 
Antares image and the images of the PSF references \mbox{$\varepsilon$~Sco}\ and Aldebaran 
in Fig.~\ref{alfsco_res2}. 
The central core of the image of \mbox{$\varepsilon$~Sco}\ (Fig.~\ref{alfsco_res2}b), 
which was observed on the same night as Antares, has a 
FWHM of 0\farcs5, which corresponds to the diffraction limit of VLT at 
17.7~\mbox{$\mu$m}.  
The peak intensity of the Antares image is 2616~Jy~arcsec$^{-2}$.  
Figure~\ref{alfsco_res1} reveals three clumpy features 
at 1\farcs5 in the north and south and 1\farcs8 in the southeast.  
Figure~\ref{alfsco_res2}a shows the same image on a different color scale, 
where these features are easier to recognize.  
The clumpy features are not present in the image of \mbox{$\varepsilon$~Sco}\ 
(Fig.~\ref{alfsco_res2}b), where up to the third Airy ring is detected. 
The upper right corner of the \mbox{$\varepsilon$~Sco}\ image 
is masked because of strong residuals of a detector 
artifact.  
However, the image quality of \mbox{$\varepsilon$~Sco}\ is much worse than Antares. 
Therefore, we checked further whether the clumpy features are real or not, 
using the image of the second PSF reference Aldebaran. 
While Aldebaran was observed on a totally different night from Antares, 
the difference between the image of \mbox{$\varepsilon$~Sco}\ and Aldebaran is 2.5\% at 
most, as shown in Fig.~\ref{alfsco_res2}d, suggesting that the PSF is 
stable.  
The image of Aldebaran (Fig.~\ref{alfsco_res2}c) shows up to the fifth Airy 
ring.  Although the fourth and fifth rings are noisy (the fourth ring is 
barely visible), there is 
no signature of the clumpy features.  Therefore, the clumpy features 
in Antares are not PSF artifacts but real.  

We checked whether the residual of 2.5\% between the PSFs from \mbox{$\varepsilon$~Sco}\ 
and Aldebaran can be explained by the difference in the orientation 
of the pupil with respect to the field of view.  However, we confirmed that 
this cannot explain the difference in two PSFs.  The residual of two PSFs 
may result from a slight difference in the telescope optics between 
the observations of \mbox{$\varepsilon$~Sco}\ and Aldebaran.

\begin{figure*}
\sidecaption
\resizebox{12cm}{!}{\rotatebox{0}{\includegraphics{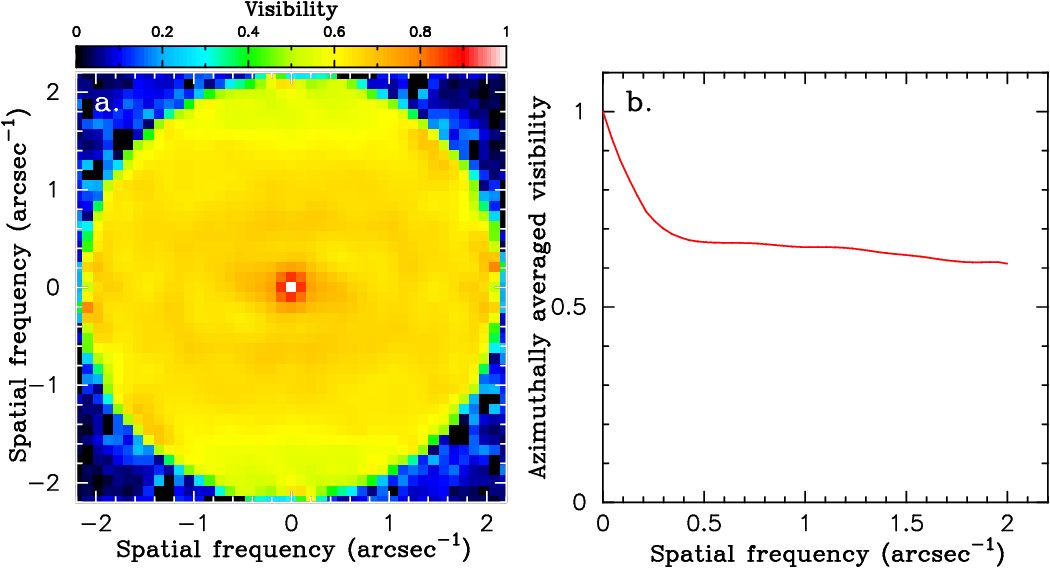}}}
%\resizebox{12cm}{!}{\rotatebox{0}{\includegraphics{alfsco_vis.ps}}}
%\resizebox{\hsize}{!}{\rotatebox{0}{\includegraphics{alfsco_vis.ps}}}
\caption{
Calibrated 17.7~\mbox{$\mu$m}\ visibility of Antares. 
{\bf a:} 2-D visibility. 
{\bf b:} Azimuthally averaged visibility. 
}
\label{alfsco_vis}
\end{figure*}

To better study the clumpy structures, we need to remove the Airy pattern 
resulting from the unresolved central star.  However, there is also 
emission from the circumstellar envelope in front of the star.  
To remove only the unresolved central star while leaving the emission 
from the circumstellar envelope intact, it is necessary to estimate the 
flux contribution of the central star at 17.7~\mbox{$\mu$m}. 
In previous studies, difference methods were taken to this end. 
For Antares, 
Marsh et al. (\cite{marsh01}) derived a maximum likelihood solution with 
the flux contribution of the central star and the (positive) background 
treated as unknowns.  As they note, this corresponds to subtracting the 
largest contribution of the central star without introducing significant 
negative residuals in the PSF-subtracted image.  
For the mid-IR imaging of Betelgeuse, 
Kervella et al. (\cite{kervella11}) estimate the flux contribution of 
the central star using the spectral energy distributions (SEDs) predicted by 
model atmospheres. 

We took a different, ``interferometric'' approach by computing the visibility, 
which is the amplitude (i.e., modulus) of the complex Fourier transform of the 
object's intensity distribution in the sky.  
We computed the Fourier transform of the images of Antares 
and the calibrator \mbox{$\varepsilon$~Sco}\ and obtained the calibrated visibility of 
Antares by dividing the (raw) visibility of Antares with that of \mbox{$\varepsilon$~Sco}. 
The calibrated 2-D visibility of Antares, shown in Fig.~\ref{alfsco_vis}a, is 
characterized by a sharp drop at low spatial frequencies (corresponding to 
the extended component) and a plateau at high spatial frequencies (corresponding 
to the unresolved component), which is clearly seen in the azimuthally averaged 
visibility (Fig.~\ref{alfsco_vis}b).  
This is typical of an object consisting of an 
unresolved central source and a well resolved extended component.  
The visibility of the plateau region corresponds to the fractional flux 
contribution of the unresolved central star.  We adopted the average of the 
visibility between a spatial frequency of 1.0 and 1.9~arcsec$^{-1}$, 0.633, 
as the fractional flux contribution of the central star.  The flux 
contribution of the central star is $1135\times0.633$ = 718~Jy.

The image of the calibrator \mbox{$\varepsilon$~Sco}\ was scaled to match this flux of 
the central star of Antares, and the flux-scaled PSF was subtracted 
from the flux-calibrated image of Antares.  
We generated a flux-scaled PSF from the Aldebaran data as well.  
For a better registration of the Antares and calibrator images, the 
images were resampled by a factor of 4 using 2-D spline interpolation 
before the subtraction, and the PSF-subtracted images were then binned back 
with four pixels.

\begin{figure*}
%\sidecaption
%\resizebox{12cm}{!}{\rotatebox{0}{\includegraphics{alfsco_psfsub.ps}}}
\resizebox{\hsize}{!}{\rotatebox{0}{\includegraphics{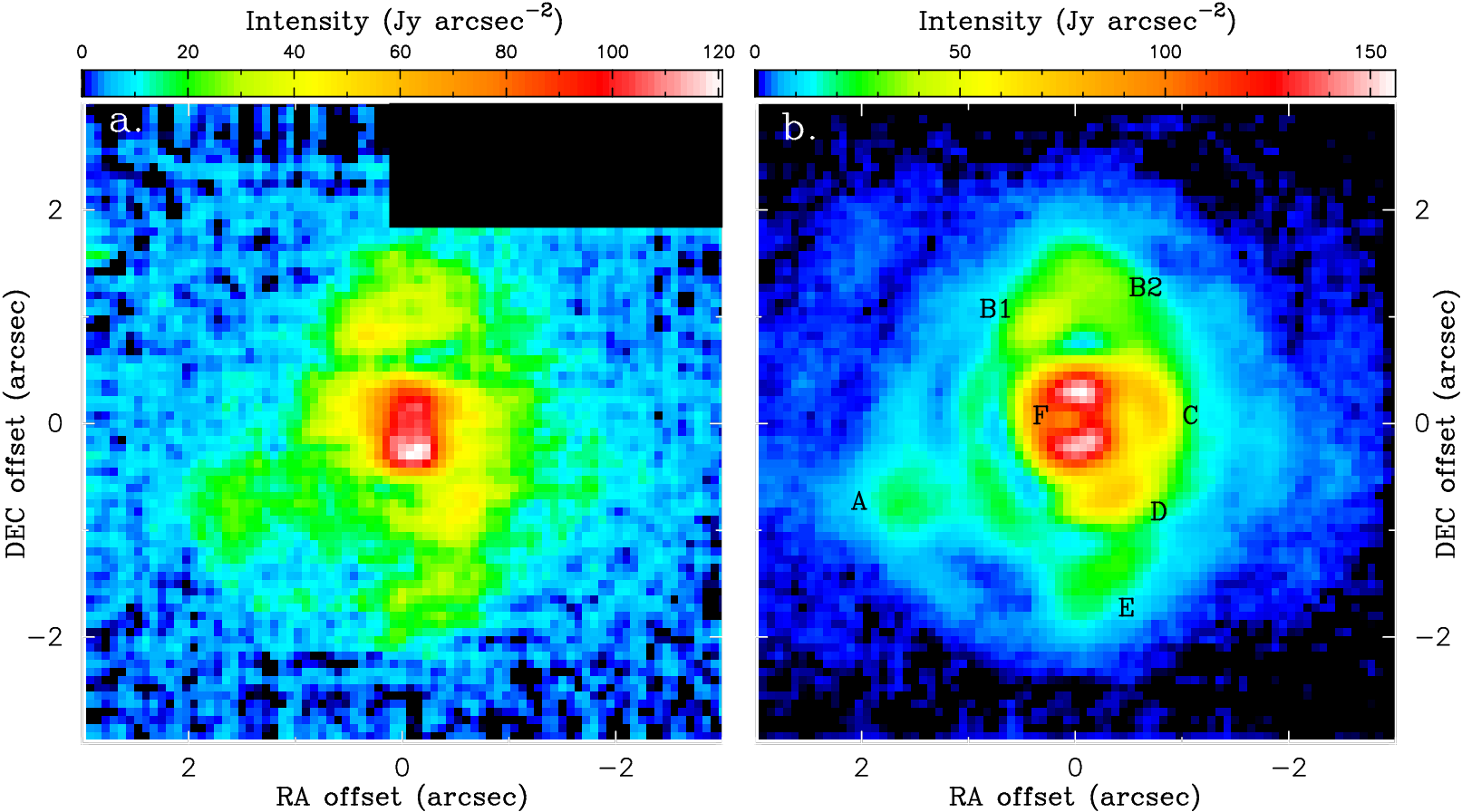}}}
%\resizebox{\hsize}{!}{\rotatebox{0}{\includegraphics{alfsco_psfsub.ps}}}
\caption{
PSF-subtracted images of Antares obtained with \mbox{$\varepsilon$~Sco}\ ({\bf a}) 
and Aldebaran ({\bf b}) as the PSF reference.  The upper right corner 
of the image in the panel {\bf a} is masked because it is affected by a 
strong detector artifact.  North is up, and east to the left. 
}
\label{alfsco_psfsub}
\end{figure*}

Figure~\ref{alfsco_psfsub} shows the PSF-subtracted image of Antares 
obtained with \mbox{$\varepsilon$~Sco}\ (Fig.~\ref{alfsco_psfsub}a) and 
Aldebaran (Fig.~\ref{alfsco_psfsub}b).  While the quality 
of the PSF-subtracted image with \mbox{$\varepsilon$~Sco}\ is not very good, three 
clumpy structures seen in Fig.~\ref{alfsco_res1} can be recognized. 
They appear more clearly in the PSF-subtracted image obtained with Aldebaran. 
The PSF-subtracted images reveal two additional clumps west and 
southwest of the star (C and D) at a distance of 0\farcs8.  
We also detected compact emission at the center with a radius of 0\farcs5. 
The central compact emission in the PSF-subtracted image with Aldebaran 
is double-peaked, while it is single-peaked in the image obtained with 
\mbox{$\varepsilon$~Sco}.  As we discuss below, this results from the uncertainty in 
the PSF, and therefore, the double peak in Fig.~\ref{alfsco_psfsub}b 
cannot be confirmed to be real.  
The radius and 
position angle of the clumps are listed in Table~\ref{clump_pos}. 
We split the northern clump B into two regions, B1 and B2, because they show 
different proper motions (with respect to the central star) 
as we discuss in the next section.

Although the PSF-subtracted images obtained with \mbox{$\varepsilon$~Sco}\ and 
Aldebaran show similar clumpy features, 
the absolute intensity in the clumps C and D, as well as the 
central compact emission F, turned out 
to differ significantly, up to 60\%.  
The reason is the aforementioned residual of 2.5\% between the PSFs obtained 
with \mbox{$\varepsilon$~Sco}\ and Aldebaran.  This residual represents the difference 
between two PSFs whose peak is normalized to 1.  When the normalized PSFs 
are scaled to the photospheric flux of Antares, even this small residual leads 
to a significant difference in the absolute intensity. 
The 2.5\% residual in the PSF is also the reason there are two peaks in the 
central compact emission in the image obtained with Aldebaran, 
while there is only a single peak in the image obtained with \mbox{$\varepsilon$~Sco}. 
It is not clear which is the better PSF---\mbox{$\varepsilon$~Sco}, which is only 9\degr\ away 
from Antares and was observed close in time but results in the noisy PSF, 
or Aldebaran, which provides a better S/N but is far away from Antares and 
was observed on a totally different night.  
Therefore, we took the mean of the intensities and the fluxes of the clumpy 
features from both PSF-subtracted images.  One half of the 
difference in the intensity and flux was adopted as the error
resulting from the uncertainty in the PSF. 
We also added the systematic error in the absolute flux calibration 
($1135\pm148$~Jy) to the uncertainties of the intensity and flux of the 
clumps.

The clumps are located at 0\farcs8--1\farcs8 from the star.  
At the distance of Antares of 170~pc, 
these angular distances correspond to 136--306~AU, which in turn 
translate into 43--96~\mbox{$R_{\star}$}\ if a linear radius of 680~\mbox{$R_{\sun}$}\ is 
adopted (Ohnaka et al. \cite{ohnaka13}).  
The radius of the compact, central emission corresponds to 
27~\mbox{$R_{\star}$}\ (= 85~AU).  
The intensity of the clumpy features A--E ranges between 23.4 and 
61.9~Jy~arcsec$^{-2}$, which is 0.9\% to 2.4\% of the peak intensity of the 
image before the PSF subtraction. 
The inner compact emission has a much higher intensity of 
135.0~Jy~arcsec$^{-2}$ (at the maximum in the south of the central star), 
which corresponds to 5.2\% of the peak intensity. 
Table~\ref{clump_pos} lists the flux integrated 
over each feature.

There are arc-like features in the PSF-subtracted image obtained with 
Aldebaran: a semi-circle going through the clump A, a smaller arc 
at $\sim$1\arcsec\ southeast of the star, and a large arc on the 
western side of the star with a radius of 1\farcs8.  
However, they are presumably residuals of the PSF subtraction.  
The image quality of the Aldebaran image is still not as good as that of 
Antares, and therefore, not all Airy rings are detected with sufficient 
S/N.  This can lead to the arc-like residuals in the 
PSF-subtracted image.

\begin{figure*}
%\sidecaption
%\resizebox{12cm}{!}{\rotatebox{0}{\includegraphics{alfsco_psfsub.ps}}}
\resizebox{\hsize}{!}{\rotatebox{0}{\includegraphics{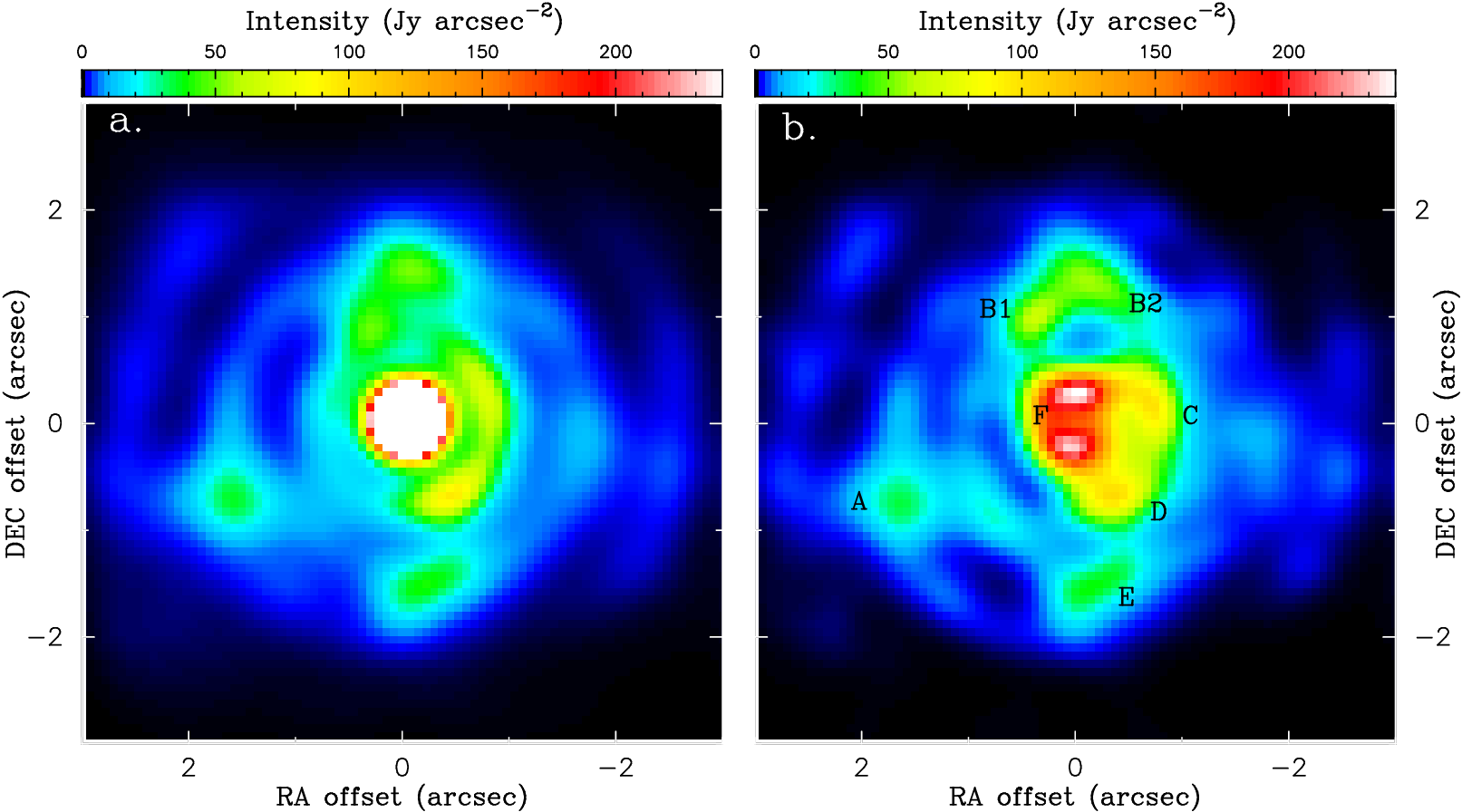}}}
%\resizebox{\hsize}{!}{\rotatebox{0}{\includegraphics{alfsco_deconv_niter5.ps}}}
\caption{
Deconvolved images of Antares obtained with the Lucy-Richardson algorithm. 
{\bf a}: Deconvolved image without the PSF subtraction.  The colors 
in the central region are saturated. 
{\bf b}: Deconvolution of the PSF-subtracted image (obtained with Aldebaran 
as the PSF).  
In both cases, the Aldebaran image was used as the PSF for the deconvolution. 
North is up, and east to the left. 
}
\label{alfsco_deconv}
\end{figure*}

We also performed the deconvolution of the Antares image to cross check 
the clumpy features seen in the PSF-subtracted images.  
We used the Lucy-Richardson algorithm (Richardson \cite{richardson72}; 
Lucy \cite{lucy74}) implemented in the STSDAS 
package of IRAF, with the Aldebaran image as the PSF.  
We stopped the deconvolution after five iterations to avoid strong artifacts.  
Figures~\ref{alfsco_deconv}a and \ref{alfsco_deconv}b show 
the deconvolved images of the original (i.e., without the PSF subtraction) 
and PSF-subtracted images of Antares, respectively.   
The figures confirm the clumps A--E seen in the PSF-subtracted images. 
The measured integrated flux of the dust clumps in the deconvolved 
images agrees with the values derived from the PSF-subtracted images.  
As discussed above, the double peak of the central emission feature F 
seen in Fig.~\ref{alfsco_deconv}b may be an artifact caused by the 
uncertainty in the PSF. 
The central emission feature F is not restored in the deconvolution of 
the non-PSF-subtracted image.  More iterations do not help restore this 
feature, either.  This is probably because the Lucy-Richardson algorithm 
tends to concentrate surrounding flux onto bright point sources, 
as Sch\"odel (\cite{schoedel10}) demonstrates.

\begin{table*}
\begin{center}
\caption {Properties of the dust clouds around Antares.  
$r$: Distance from the central star in units of arcseconds and stellar radii. 
PA: Position angle. 
$I_{\rm peak}$: Peak intensity.
Flux: Flux integrated over each cloud.
$T_{\rm d}$: Dust temperature. 
$M_{\rm d}$: Dust mass. 
$r$(1998): Distance from the central star in the 20.8~\mbox{$\mu$m}\ image of 
Marsh et al. (\cite{marsh01}) taken in 1998. 
$\Delta r$: Angular displacement between 1998 and 2010. 
$V$: Velocity of the outward motion projected onto the plane of the sky.  
$\dagger$: The distance for the clump F actually represents the radius of 
the inner, compact emission. 
}

\begin{tabular}{l l l r r r l l l l l}\hline
ID & $r$ & $r$ & PA & $I_{\rm peak}$ & Flux & $T_{\rm d}$ & $M_{\rm d}$ & $r$(1998)
& $\Delta r$ & $V$ \\
  & (\arcsec)&(\mbox{$R_{\star}$})&(\degr) & (Jy~arcsec$^{-2}$) & (Jy) & (K) & (\mbox{$M_{\sun}$}) &
(\arcsec) & (\arcsec) & (\mbox{km s$^{-1}$}) \\
\hline
A & 1.8 & 96 & 115 & $23.4\pm3.9$ & $14.0\pm2.3$ & 280 & $5\times10^{-9}$ & 1.3 & $0.5\pm0.1$ & $34\pm7$ \\
B1 & 1.0 & 53 & 17 & $50.8\pm7.8$ & $16.5\pm2.2$ & 370 & $3\times10^{-9}$ & 
0.4 & $0.6\pm0.1$ & $40\pm7$ \\
B2 & 1.3 & 69 & 353 & $36.7\pm5.4$ & $25.0\pm3.7$ & 320 & $5\times10^{-9}$ &
1.0 & $0.2\pm0.1$ & $13\pm7$ \\
C & 0.8 & 40 & 289 & $61.9\pm22.2$ & $27.7\pm10.9$ & 430& $3\times10^{-9}$ &
--- & --- & --- \\
D & 0.8 & 40 & 213 & $60.8\pm21.0$ & $29.5\pm9.5$ & 430& $3\times10^{-9}$ &
--- & --- & --- \\
E & 1.5 & 78 & 189 & $28.9\pm5.3$ & $16.0\pm2.4$ & 310& $4\times10^{-9}$ & 1.0 & $0.5\pm0.1$ & $34\pm7$ \\
F & 0.5$^{\dagger}$&27$^{\dagger}$&---& $135.0\pm31.8$ &$87.6\pm30.2$&550&$6\times10^{-9}$&---&---&---\\
\hline
\label{clump_pos}

\end{tabular}
\end{center}
\end{table*}

\section{Discussion}
\label{sect_discuss}

\subsection{Physical properties of dust clouds}
\label{subsect_prop}

The modeling of the mid-IR spectrum and interferometric data of Antares 
by Danchi et al. (\cite{danchi94}) and the SED modeling 
by Verhoelst et al. (\cite{verhoelst09}) suggest that the 
17.7~\mbox{$\mu$m}\ flux is dominated by dust emission.  
Harper et al. (\cite{harper09}) report the detection of \mbox{[Fe II]} lines at 
17.94 and 24.52~\mbox{$\mu$m}\ in a sample of RSGs including Antares. 
The former emission line is included in the wavelength range covered by the 
Q1 filter.  Antares was observed only for the 24.52 [Fe II] line and 
not for the [Fe II] line at 17.94~\mbox{$\mu$m}.  
However, Harper et al. (\cite{harper09}) show that the 17.94~\mbox{$\mu$m}\ 
[Fe II] line forms close to the star at $\sim$1.5~\mbox{$R_{\star}$}, which is unresolved 
with the spatial resolution of VISIR at 17.7~\mbox{$\mu$m}.  
Therefore, the contribution 
of the [Fe II] line to the clumpy features is likely to be negligible.  

The circumstellar dust around Antares is optically thin.  
For example, Danchi et al. (\cite{danchi94}) modeled 
the mid-IR spectrum (7--23~\mbox{$\mu$m}) and the 11~\mbox{$\mu$m}\ visibility 
observed for Antares using a spherical shell and derived an 11~\mbox{$\mu$m}\ 
optical depth of 0.011.  
We computed the temperature distribution in an optically thin, spherical dust 
shell with $\tau_{11\mbox{$\mu$m}}=0.011$ using our Monte Carlo code presented 
in Ohnaka et al. (\cite{ohnaka06}).  
While the circumstellar envelope of Antares is obviously not spherical, 
the dust temperature computed in spherical symmetry is a good approximation 
for an optically thin case. 
We used the optical properties of silicate presented by Draine \& Lee 
(\cite{draine84}) and a grain size distribution characterized by an 
exponent of $-3.5$ between 0.005 and 0.25~\mbox{$\mu$m}. 
With these grain properties, the 11~\mbox{$\mu$m}\ optical depth of 0.011 
translates into an optical depth of 0.11 at 0.55~\mbox{$\mu$m}.  
The central star was approximated by the blackbody of 3700~K, adopting 
the effective temperature determined by Ohnaka et al. (\cite{ohnaka13}).   
The density was assumed to decrease as $r^{-2}$, and 
we estimated the inner radius of the dust shell from the 12.5~\mbox{$\mu$m}\ image 
of Antares taken by Marsh et al. (\cite{marsh01}).  They found a ring with 
a radius of 0\farcs3, which corresponds to 16~\mbox{$R_{\star}$}\ with the stellar angular 
radius of 18.5~mas measured by Ohnaka et al. (\cite{ohnaka13}).  
The outer radius of the shell was set to $1.6\times10^5$~\mbox{$R_{\star}$}.

The dust temperatures derived at the position of the clumps from this 
model range from 280 to 430~K, as listed in Table~\ref{clump_pos}.  
The dust temperature at the outer edge of the inner and compact emission F 
(27~\mbox{$R_{\star}$}) is 550~K  
(dust temperature predicted at the inner radius is 780~K).  
We note that the distance of the individual dust clouds from the star 
is projected onto the plane of the sky, and therefore the true 
radial distance from the star can be larger than derived from the VISIR 
image.  This means that the temperatures of the dust clouds derived above 
are upper limits.  However, given that the emission from the dust clouds 
is significant at 17.7~\mbox{$\mu$m}, the peak of the blackbody 
radiation from the dust clouds should be shortward of 17.7~\mbox{$\mu$m}.  
This means that the temperatures of the dust clouds would not be lower than 
$\sim$170~K.  Multicolor mid-IR imaging would be useful for reliably 
determining the temperature of the dust clouds.

We estimated the dust mass ($M_{\rm d}$) of the clumps from the dust 
temperature ($T_{\rm d}$) and the flux integrated over each clump ($F_{\nu}$) 
by 
\[
M_{\rm d} = F_{\nu} d^{2}/(\kappa_{\nu} B_{\nu}(T_{\rm d})), 
\]
where $d$ is the distance to Antares, $B_{\nu}(T_{\rm d})$ is the Planck 
function, and 
$\kappa_{\nu}$ the mass absorption coefficient of the dust in units of 
cm$^{2}$~g$^{-1}$.  Using the absorption cross section of the astronomical 
silicate presented by Draine \& Lee (\cite{draine84}) 
and adopting a grain bulk density of 3~g~cm$^{-3}$, we obtain 
$\kappa_{\nu} \approx 1000$~cm$^{2}$~g$^{-1}$ at 17.7~\mbox{$\mu$m}.  
The estimated dust mass of the clumps ranges from $3\times10^{-9}$ 
to $6\times10^{-9}$~\mbox{$M_{\sun}$}, as listed in Table~\ref{clump_pos}.  
Given the aforementioned uncertainty in the dust temperature, the 
uncertainty in the dust mass amounts to a factor of 2.  
If we adopt the mass-loss rate of 
$2\times10^{-6}$~\mbox{$M_{\sun}$~yr$^{-1}$}\ (Braun et al. \cite{braun12}) and a gas-to-dust 
ratio of 100 (Draine \& Li \cite{draine07}), 
the amount of dust ejected in one year is $2\times10^{-8}$~\mbox{$M_{\sun}$~yr$^{-1}$}.  
This is greater than the derived dust mass of the clumpy clouds by a factor 
of 3--7, implying the presence of continuous mass loss superimposed 
on the clumps.  
The extended circumstellar envelope (apart from the clumps), which is 
clearly seen in the azimuthally averaged intensity profile 
(Fig.~\ref{alfsco_img1Davg}), corresponds to this continuous mass loss.

\begin{figure*}
%\sidecaption
%\resizebox{12cm}{!}{\rotatebox{0}{\includegraphics{alfsco_psfsub.ps}}}
\resizebox{\hsize}{!}{\rotatebox{0}{\includegraphics{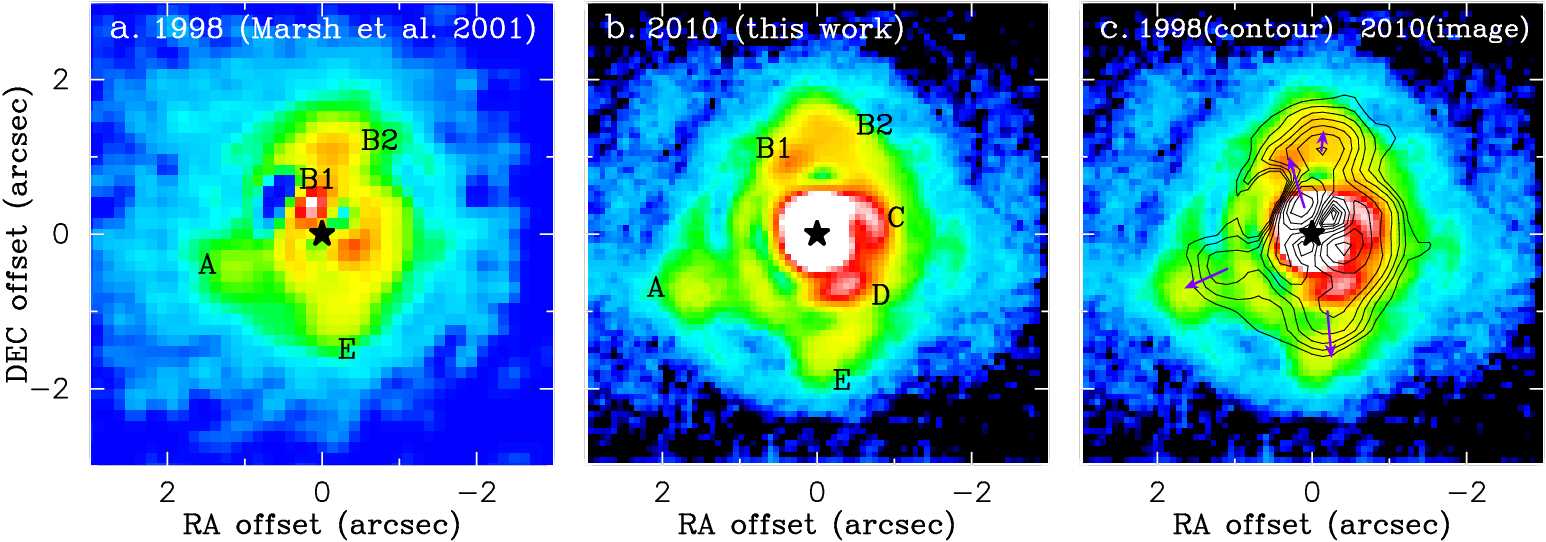}}}
%\resizebox{\hsize}{!}{\rotatebox{0}{\includegraphics{alfsco_timevar.ps}}}
\caption{
Temporal evolution of the dust clouds between 1998 and 2010. 
{\bf a:} PSF-subtracted image of Antares taken at 20.8~\mbox{$\mu$m}\ in 
1998 by Marsh et al. (\cite{marsh01}). 
{\bf b:} PSF-subtracted image obtained in 2010 in this work using Aldebaran 
as the PSF reference.  The color in the central core is saturated to show the 
clumpy structures clearly. 
{\bf c:} Comparison between the PSF-subtracted images taken in 1998 
(contour) and 2010 (image).  
The outward motions of the clumps A, B1, B2, and E are marked by the arrows. 
The position of the central star is marked with $\star$ in all panels. 
North is up, and east to the left. 
}
\label{alfsco_timevar}
\end{figure*}

\subsection{Detection of the outward motions of dust clouds}
\label{subsect_motions}

Marsh et al. (\cite{marsh01}) obtained 12.5 and 20.8~\mbox{$\mu$m}\ images of 
Antares with the Keck telescope in June 1998, 12 years before our VISIR 
observation, with a spatial resolution of 0\farcs48, which is 
almost the same as our VISIR image.  
They detected a ring with a radius of $\sim$0\farcs3 
and three discrete clumps at a radius of $\sim$1\farcs2 from the star.  
We compared their 20.8~\mbox{$\mu$m}\ image and our 17.7~\mbox{$\mu$m}\ VISIR image 
to see possible time variations in the clumpy structure of the circumstellar 
envelope.  Figure~\ref{alfsco_timevar} shows the PSF-subtracted image 
read off Fig.~1 of Marsh et al. (\cite{marsh01})\footnote{Unfortunately, 
the original electronic data are no longer available 
(K.~A.~Marsh, priv. comm.).} 
and our VISIR image.  The figure reveals that the overall structure of the 
envelope is more extended in our VISIR image than in the Keck image. 
The wavelengths of the 1998 Keck image and our VISIR image are slightly 
different (20.8~\mbox{$\mu$m}\ and 17.7~\mbox{$\mu$m}, respectively).  
However, because their 
wavelength is longer than ours, their image samples colder regions, which are 
farther away from the star.  In other words, if an image had been 
taken at 17.7~\mbox{$\mu$m}\ in 1998, the ring and the clumpy features would have 
appeared closer to the star.  Therefore, the difference in wavelength 
is unlikely to be responsible for the time variation of the 
clumpy features. 
The inner compact emission is not visible in the Keck image, which 
suggests that this emission component is newly formed dust ejected 
after 1998. 

Figure~\ref{alfsco_timevar} allows us to trace the expansion of the individual 
clumps.  As Table~\ref{clump_pos} summarizes, the dust clumps A, B1, and E 
have moved over 0\farcs5--0\farcs6 and the clump B2 over 0\farcs2 in 12 years.  
We estimate the uncertainty 
in the angular displacement to be $\pm$0\farcs1, which results from a 
half of the pixel size of the 1998 Keck image (0\farcs138) and our 2010 
VISIR image (0\farcs075).  At the distance of 170~pc, 
the proper motions of 0\farcs5--0\farcs6 and 0\farcs2 correspond to distances 
of $(1.3-1.5)\times10^{15}$~cm (= $85-102$~AU = $27-32$~\mbox{$R_{\star}$}) and 
$5.1\times10^{14}$~cm (= 34~AU = 11~\mbox{$R_{\star}$}), respectively, projected 
onto the plane of the sky. 
These displacements in 12 years translate into velocities of 34, 40, 13, and 
34~\mbox{km s$^{-1}$}\ for the clumps A, B1, B2, and E, respectively, with an 
uncertainty of $\pm7$~\mbox{km s$^{-1}$}\ (see Table~\ref{clump_pos}).  
Because we do not know the 3-D positions of the dust clouds, the radial 
expansion velocity with respect to the star is even 
higher.  On the other hand, we cannot recognize clear outward motions 
for the clumps C and D.  While these clumps may be moving more in parallel 
to the line of sight, 
we note that the morphology of the clumps C and D has changed.  
They appeared as a single clump in 1998, but they clearly appear as two 
distinct clumps in 2010. 
Because the spatial resolution of the 1998 Keck image is almost the same 
as our VISIR image, the difference in the appearance of C and D 
cannot be attributed to the insufficient 
spatial resolution of the 1998 image.  Perhaps 
the large clump at $\sim$0\farcs5 in the west to southwest 
seen in the 1998 image 
may have moved outward, leading to much weaker emission in 2010, and 
the clumps C and D may be newly formed dust clouds ejected after 1998.

The measured velocity of the clumps A, B1, and E is remarkably higher than the 
expansion velocity of $17.3\pm3.4$~\mbox{km s$^{-1}$}\ and $\sim$20~\mbox{km s$^{-1}$}\ derived by 
Bernat (\cite{bernat77}) and Braun et al. (\cite{braun12}), respectively. 
This implies that the individual clouds may be ejected from the star at 
different velocities. 
Baade \& Reimers (\cite{baade07}) detected absorption lines blueshifted 
by 0.5--19.9~\mbox{km s$^{-1}$}\ with respect to the star in the UV spectra of Antares, 
which suggests episodic and/or clumpy mass loss.  
The analysis of the CO fundamental lines near 4.6~\mbox{$\mu$m}\ 
in a sample of red giants and supergiants by Bernat (\cite{bernat81}) 
also shows multiple components expanding at different velocities.  
It cannot be explained simply by the acceleration of material, because 
there is no correlation between the temperature and expansion velocity 
of the different components: the faster components do not necessarily 
show lower temperatures (resulting from being located farther out), 
as expected from the simple acceleration.  
These results imply the random nature of the dust cloud ejection 
mechanism.  For example, Wittkowski et al. (\cite{wittkowski11}) 
propose that the inhomogeneities in the outer atmosphere of 
Mira stars might be caused by pulsation- and shock-induced chaotic 
motions.  Such chaotic motions may be possible in RSGs as well, 
although the pulsation 
amplitude is much smaller than Mira stars.  Convective motions and/or 
magnetohydrodynamical (MHD) processes in 
the photosphere might also be responsible for the random nature of 
the ejection velocity of the dust clouds.

The absence of the inner emission component F 
in the Keck image taken in 1998 suggests new dust formation after 1998.  
If the dust formed in 1998 reached the radius of 0\farcs5 in 2010, the 
expansion velocity is estimated to be 34~\mbox{km s$^{-1}$}, which agrees with the 
velocities of the dust clouds A and E derived above.  
However, given that the expansion velocity can be at least as high as 40~\mbox{km s$^{-1}$}\ 
as in the clump B1, it is 
possible that the dust formation responsible for the inner component F 
occurred more recently.  

The presence and absence of dust in the inner region of the envelope around 
Antares has been discussed in previous studies.  
Based on the 11~\mbox{$\mu$m}\ interferometric data, Sutton et al (\cite{sutton77}) 
showed that dust exists farther than a radius of 0\farcs2 from the star.  
The 11~\mbox{$\mu$m}\ imaging of Bloemhof \& Danen (\cite{bloemhof95}) confirms 
the absence of dust within a radius of $\sim$0\farcs6 from the star 
for the data taken in 1985.  
The PSF-subtracted image obtained at 11~\mbox{$\mu$m}\ in 1991 by 
Danchi et al. (\cite{danchi92}) also shows that the 
intensity at the stellar position is lower than in the surrounding region, 
suggesting a lower amount of dust near the star. 
The 11~\mbox{$\mu$m}\ interferometric observations and modeling by 
Danchi et al. (\cite{danchi94}) suggest an inner radius of the dust 
envelope of $\sim$1\arcsec. 
The mid-IR images taken in 1998 by 
Marsh et al. (\cite{marsh01}) show a ring of dust with a radius of 0\farcs3. 
They estimated that this ring had been ejected 10 to 20 years before their 
observation in 1998, with the region around the star's position 
devoid of dust.  
Then our VISIR image suggests the ejection 
of newly formed dust after 1998.  These results confirm that the dust 
formation takes place in an episodic manner.

Similar clumpy structures and/or episodic mass loss are suggested for 
other RSGs as well.  The mid-IR imaging of Betelgeuse by 
Kervella et al. (\cite{kervella11}) reveals clumpy structures extending to 
$\sim$2\arcsec (= 95~\mbox{$R_{\star}$}) from the star.  
Based on the interferometric measurements at 
11~\mbox{$\mu$m}, Tatebe et al. (\cite{tatebe07}) suggest an episodic ejection 
of material at 75~\mbox{km s$^{-1}$}\ in \object{$\alpha$~Her}.  
Danchi et al. (\cite{danchi01}) 
detected the outward motion of two discrete shells in 
the dust-enshrouded RSG NML~Cyg.  
Therefore, the mass loss from RSGs---whether 
optically bright or very dusty---may intrinsically show episodic and 
clumpy features in addition to continuous outflows.

\section{Concluding remarks}
\label{sect_concl}

We obtained a diffraction-limited image of the RSG Antares 
at 17.7~\mbox{$\mu$m}\ with a spatial resolution of 0\farcs5.  The image 
shows six clumpy dust clouds at 0\farcs8--1\farcs8 away from the star, 
as well as the inner compact emission with a radius of 0\farcs5. 
The clumpy dust clouds have peak intensities of 0.9\% to 2.4\% of the 
central star, while the inner emission has a much higher intensity of 
5.2\% of the central star.  We set upper limits of 280--550~K on the 
temperature in the dust clouds.  The estimated dust mass of the individual 
clouds is $(3-6)\times10^{-9}$~\mbox{$M_{\sun}$}, which is less than the 
amount of dust ejected in one year by a factor of 3--7.  
This means that the mass loss from Antares consists of the ejection of 
clumpy dust clouds and a continuous outflow.

We detected the outward motions of the dust clouds by comparing with 
the 20.8~\mbox{$\mu$m}\ image taken in 1998 with the Keck telescope.  The 
measured proper motions amount to 0\farcs2--0\farcs6 in 12 years.  
The expansion velocity of the dust clouds (projected onto the plane of 
the sky) is 13--40~\mbox{km s$^{-1}$}.  The inner compact emission is not 
seen in the Keck image taken in 1998, suggesting that newly formed dust 
was ejected between 1998 and 2010.  
If we adopt the radial expansion velocity of 34~\mbox{km s$^{-1}$}\ 
for the inner component, it must have been ejected in 1998.  
However, given that we derived a velocity as high as 40~\mbox{km s$^{-1}$}\ in one 
cloud, it is also possible that the new dust formation occurred 
later than 1998.  
Spatially resolved spectroscopy of the individual dust clouds would be 
useful for probing their kinematics and obtaining a 3-D picture of the 
clumpy circumstellar envelope. 
The outward velocities of the dust clumps cannot be explained by a simple 
accelerating outflow, implying the random nature of the dust cloud 
ejection mechanism.

\begin{acknowledgement}
We thank the ESO VLT team for supporting our VISIR observations.  
This research made use of the SIMBAD database,
operated at the CDS, Strasbourg, France.  
\end{acknowledgement}

\end{document}